# Improved in-car sound pick-up using multichannel Wiener filter

Juhi Khalid
School of Electrical Engineering
and Computer Science
University of Ottawa
Ottawa, Canada
jkhal027@uottawa.ca

Martin Bouchard
School of Electrical Engineering
and Computer Science
University of Ottawa
Ottawa, Canada
bouchm@uottawa.ca

*Abstract*—With advancements in automotive electronics and sensors, the sound pick-up using multiple microphones has become feasible for hands-free telephony and voice command in-car applications. However, challenges remain in effectively processing multiple microphone signals due to bandwidth or processing limitations. This work explores the use of the Multichannel Wiener Filter algorithm with a two-microphone in-car system, to enhance speech quality for driver and passenger voice, i.e., to mitigate notch-filtering effects caused by echoes and improve background noise reduction. We evaluate its performance under various noise conditions using modern objective metrics like Deep Noise Suppression Mean Opinion Score. The effect of head movements of driver/passenger is also investigated. The proposed method is shown to provide significant improvements over a simple mixing of microphone signals.

## I. Introduction

With advancements in automotive electronics and sensors, sound pick-up using multiple microphones has become feasible for hands-free telephony and voice command in-car applications. For example, in addition to a primary microphone mainly dedicated to sound pick-up of the driver voice, a secondary microphone can be added closer to the front seat passenger to improve sound pick-up for the passenger voice. But due to bandwidth or processing limitations, it may still not be possible to fully process or transmit the two microphone signals separately. Therefore, some solutions need to be developed, such as mixing the microphone signals or switching between the two microphone signals. However, each of these solutions comes with drawbacks.

Switching between the microphone signals can be done based on which speech source is active (driver or passenger). The main drawback of this approach is that it creates sudden changes in the acoustic echo path between loudspeakers and the selected microphone to use (primary for driver or secondary for passenger), which will affect the performance of acoustic echo cancellation (AEC) systems. AEC systems are a requirement to prevent that a voice ("far-end") played in car loudspeakers be fed back to the microphones before transmission or processing. Acoustic echo cancellation is also used during media playback to help with the voice command functionality. A second drawback of microphone switching is its inability to properly deal with the case of the driver and passenger speaking simultaneously.

Mixing the microphone signals has its own drawbacks. Since each speech source reaches the summed microphone signal through two acoustic paths (one for each microphone), the sum of two signals with different delays can lead to an echo in the resulting time signal and a "notch filtering" effect in the spectrum of the resulting signal, i.e., distortion because some frequencies are attenuated at the notches. The background noise power is also increased (typically doubled) when two microphone signals are added in mixing.

While the use of the multichannel Wiener filter (MWF) for speech enhancement has been considered before in car environments, it has been either for solely removing background noise from a single talker [1]-[3], for positioning virtual acoustic sources during conference calls performed in cars [4], or in the context of SONAR-like driver assistance [5]. In this paper, we propose the use of MWF adaptive filtering for the task of notch-filtering effect mitigation (echo removal) in the sum of microphone signals and for the task of multichannel background noise reduction. It should be noted that these tasks can be simultaneously achieved in practice, and that they can be performed with either one or two speakers active (or multiple speakers, if additional microphones are used). We evaluate the MWF performance under various noise conditions. The effect of head movements of driver/passenger is also investigated.

As detailed in the next section, the MWF performs source extraction for each voice source (i.e., each source extracted with a distinct MWF filter, with background noise attenuated). This removes the multi-path speech propagation and echo in the speech from each source. Then the extracted sources can be added (mixed) before further transmission or processing. This has direct use cases for hands-free telephony, as well as voice command recognition (with or without media playing). If MWF is to be enabled while media is playing, since the loudspeakers increase the number of directional sources that the MWF needs to deal with, and since the number of sources should not exceed the number of microphones for MWF to work effectively, AEC can be used to remove the effect of the played media on the signals picked up by the microphones, at the cost of higher complexity. Since the MWF extracts the speech sources before combining them, it can also be used for interference cancellation (i.e., cancelling competing talkers if more than one voice is active).

The outline of the paper is as follows. The next section presents the mathematical formulation of the MWF method.

The methodology used to test our proposed method is presented in Section III. Experimental results are provided in Section IV, while some discussion and a conclusion are provided in the last sections.

## II. MULTICHANNEL WIENER FILTER (MWF)

The traditional Wiener filter is a signal processing technique which can be used for noise reduction and signal enhancement in speech processing. It operates on a single-channel input and aims to estimate a desired signal (e.g., clean speech signal) by minimizing the mean square error between the estimated and actual signals. It relies on the knowledge of the target signal (clean speech) statistics and background noise statistics, to create an optimal filter that enhances the target signal while suppressing noise. The MWF extends this concept to multiple input signals (multi-microphones). Instead of processing a single noisy signal, MWF exploits the spatial and statistical correlations between multiple microphone signals to extract the desired speech while suppressing background noise and interference. This makes it particularly useful in scenarios with multiple competing sources. Since the statistics are continuously estimated and updated, the MWF is also suitable for dynamic environments.

The MWF has some equivalences with beamforming algorithms [6], where the term beamforming here is not restricted to a microphone array of closely located microphones, i.e., microphones can be distributed in a car (e.g., the primary and secondary microphones for driver and front seat passenger). Beamformers are typically designed based on knowing a propagation model for a sound coming from the direction of a target speech source, as well as knowing the multichannel (multi-microphone) correlation matrix of the signals to be minimized (i.e., background noise and other competing speech sources). The correlation matrix can also be estimated based on propagation models for the noise and competing sources. Beamformers may thus require knowing the direction of arrival of the target and competing sources. Instead of relying on a propagation model and direction of arrival for the sources, the MWF formulation is entirely based on estimating correlations between the different microphone signals and the target speaker signal to extract (e.g., driver voice). The resulting correlations are used to build a multichannel correlation matrix and multichannel correlation vector, from which the MWF solution is computed. The required correlations can be estimated if it is possible to detect when the different sources are active, separately or jointly. Therefore, no acoustic propagation model or direction of arrival estimation is required in the MWF approach.

The use of the MWF algorithm has been suggested in several applications (e.g., for hearing aids [7]). But using the MWF can be difficult for applications where it is not simple to determine when each individual source is active, or where the condition of having single sources active may not be frequent. However, in the car environment, with one microphone mostly dedicated to each source (driver, passenger), the detection of the active source is easier to perform and single talker scenarios are frequent, allowing to perform the estimation of the required correlations in the MWF filter.

Consider the driver and passenger speech source signals, $s_A$ and $s_B$, with background noise signal $v_i$ at the $i^{th}$ microphone. $s_{iA}$ and $s_{iB}$ are the signals received at the $i^{th}$ microphone from source $s_A$ and $s_B$, respectively. Then the signals received at microphones 1 and 2 can be written as:

$$x_1(n) = s_{1A}(n) + s_{1B}(n) + v_1(n) \quad (1)$$

$$x_2(n) = s_{2A}(n) + s_{2B}(n) + v_2(n) \quad (2)$$

Where the sources are uncorrelated to each other as well as with the noise, i.e.,

$$\mathrm{E}[s_{jA}(n)s_{iB}(n)] = 0; \quad i,j = 1,2 \quad (3)$$

$$\mathrm{E}[s_{jk}(n)v_i(n)] = 0; \quad i,j = 1,2 \quad k = A,B \quad (4)$$

Since frequency domain implementations allow a more efficient processing and a simpler formulation, the solution of the MWF will be described below in the frequency domain, where the frequency content of signals is assumed to be computed on a frame-by-frame basis (e.g., with FFTs). The speech sources to be estimated are $\hat{s}_{1A}$ from the driver at mic. 1 (primary) and $\hat{s}_{2B}$ from the passenger at mic. 2 (secondary). Each source is estimated from a different MWF. Without loss of generality, we provide below the equations obtain the estimated source $\hat{s}_{1A}$ from the driver at mic. 1:

$$\hat{s}_{1A}(\omega) = \mathbf{w}^H(\omega)\mathbf{x}(\omega) = [w_1(\omega)\,w_2(\omega)][x_1(\omega)\,x_2(\omega)]^T \quad (5)$$

Where

$$\mathbf{w}(\omega) = \mathbf{R}^{-1}(\omega)\mathbf{p}(\omega) \quad (6)$$

$$\mathbf{w}(\omega) = [w_1^*(\omega)\,w_2^*(\omega)]^T \quad (7)$$

$$\mathbf{R}(\omega) = \begin{bmatrix} \Phi_{x_1x_1}(\omega) & \Phi_{x_1x_2}(\omega) \\ \Phi_{x_2x_1}(\omega) & \Phi_{x_2x_2}(\omega) \end{bmatrix} = \begin{bmatrix} \Phi_{x_1x_1}(\omega) & \Phi_{x_1x_2}(\omega) \\ \Phi_{x_1x_2}^*(\omega) & \Phi_{x_2x_2}(\omega) \end{bmatrix} \quad (8)$$

$$\mathbf{p}(\omega) = [\Phi_{x_1d}(\omega) \quad \Phi_{x_1}(\omega)]^T = [\Phi_{dx_1}^*(\omega) \quad \Phi_{dx_2}^*(\omega)]^T \quad (9)$$

with $\Phi_{yx}(\omega) = \mathrm{E}[y(\omega)x^*(\omega)]$ and $d = s_{1A}$ to extract.

The elements of the **R** matrix are given by,

$$\Phi_{x_jx_i}(\omega) = \Phi_{s_{jA}s_{iA}}(\omega) + \Phi_{s_{jB}s_{iB}}(\omega) + \Phi_{v_jv_i}(\omega) \quad (10)$$

$\Phi_{x_jx_i}(\omega)$ can be estimated as a sum of components when only single sources are active (i.e., only driver or passenger voice, with noise) and then subtracting noise-only statistics:

$$\Phi_{x_jx_i}(\omega) = \left(\Phi_{s_{jA}s_{iA}}(\omega) + \Phi_{v_jv_i}(\omega)\right)$$
$$+ \left(\Phi_{s_{jB}s_{iB}}(\omega) + \Phi_{v_jv_i}(\omega)\right) - \Phi_{v_jv_i}(\omega) \quad (11)$$

The elements of **p** are given by:

$$\Phi_{x_jd}(\omega) = \Phi_{s_{jA}s_{1A}}(\omega) \quad (12)$$

$\Phi_{x_jd}(\omega)$ can be estimated when only the source $s_A$ is active (i.e., driver with noise) and subtracting the noise-only statistics:

$$\Phi_{x_jd}(\omega) = \Phi_{s_{jA}s_{1A}}(\omega)$$
$$= \left(\Phi_{s_{jA}s_{1A}}(\omega) + \Phi_{v_jv_i}(\omega)\right) - \Phi_{v_jv_i}(\omega) \quad (13)$$

In order to adapt to time-varying speech and noise statistics and time-varying source locations, in practice all correlations are calculated as a moving average with a forgetting factor of $\lambda$:

$$\Phi_{yx}(\omega,m) = \lambda\Phi_{yx}(\omega,m-1) + (1-\lambda)\Phi_{inst.yx}(\omega,m) \quad (14).$$

Where $m$ is the index of the current signal frame being processed, and $\Phi_{inst.yx}(\omega,m)$ is the instantaneous estimate obtained from the current frame. While calculating the inverse of the correlation matrix in (6), to prevent poor performance and too much noise amplification when the matrix is poorly conditioned, some regularization of the matrix is performed as below, where $\delta$ is a normalized regularization factor:

$$\mathbf{w} = \left(\mathbf{R} + \delta \frac{\operatorname{tr}(\mathbf{R})}{\operatorname{size}(\mathbf{R})}\mathbf{I}\right)^{-1} \quad (15).$$

If the extracted voice sources are to be combined before transmission or further processing, the extracted driver speech at mic. 1 $\hat{s}_{1A}$ and the extracted passenger speech at mic. 2 $\hat{s}_{2B}$ can simply be added, without notch filtering effect and with reduced background noise (and regardless of whether only one or two of these sources are active at any given time).

### III. METHODOLOGY

The adaptive MWF approach is implemented to address the problem of two-path propagation and the resulting notch filtering effect for in-vehicle audio systems when combining microphone signals, which also results in increased background noise. It is implemented in the frequency domain. The MWF is applied using the Overlap-Add (OLA) method with 50% overlap, ensuring computational efficiency. The method derives statistical computations directly from the microphone signals. The MWF processing is placed right after the microphone sound pick-ups, and the resulting output signals can be sent for further processing such as acoustic echo cancellation (AEC). This is shown in Fig. 1.

**Active Speaker Detection:** This can be done various ways, including:

- Comparing instantaneous or smoothed power of microphone signals
- Comparing correlations between microphone signals and past MWF output signal samples

For the experiments performed in the rest of this paper, in order to decouple the effect of active speaker detection from the performance achievable by MWF filtering, a perfect classification assumption is made, i.e., perfect knowledge of when each of the driver and passenger voice source is active.

**Microphone setup:** The impulse responses of the vehicle cabin are simulated using the RIR Generator [8], which uses the classic image method [9] and simulates a rectangular enclosure. The simulation is done based on the following assumptions:
- Car dimensions: 5 × 2 × 1.78 meters
- Reverberation time: 0.07 seconds
- Cardioid microphones placed 0.8 meters apart, with the driver's microphone as primary and the passenger's microphone as secondary, each pointing towards the respective speaker.

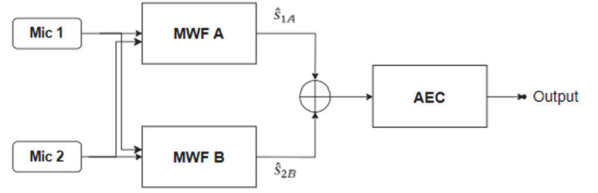

Fig. 1. Position of MWF in typical setup.

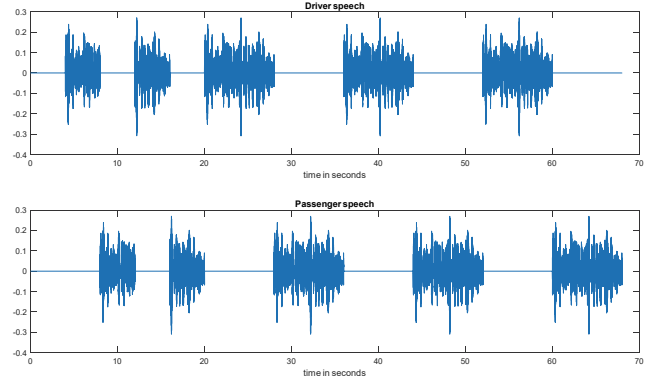

Fig. 2. Speech signals used for testing.

TABLE I. MICROPHONE AND SOURCE POSITIONS IN THE SIMULATED CAR

| Microphone/Source | Distance from the front (m) | Distance from the left side (m) | Distance from the floor (m) |
|---|---|---|---|
| Primary Microphone | 1.65 | 0.6 | 1.7 |
| Secondary Microphone | 1.65 | 1.4 | 1.7 |
| Driver | 2.5 | 0.6 | 0.75 |
| Passenger | 2.5 | 1.4 | 0.75 |

- The resulting impulse responses were symmetric between the two sources and the two microphones, and the cross-side impulse responses had a gain 2 dB weaker than the same-side impulse responses.

The exact positions of the microphones, driver and passenger used for simulation are given in Table I.

**Speech signals:** Intermittent speech is used for the driver and the passenger, and there is no overlap between speech segments in the experiments. The sampling rate is set to 16 kHz, as wideband speech is commonly used for voice communication systems. The driver and passenger speech signals used for testing are shown in Fig. 2. The same speech sequence was used for both driver and passenger speech, to reduce the variations in some metrics that depend on speech content.

**Noise types:** Multiple types of spatially uncorrelated noises are used to evaluate the noise suppression performance of MWF: white noise, red noise, pink noise, green noise, and hoth noise. White noise has a flat power spectrum, whereas red and pink noises follow $\frac{1}{f^2}$ and $\frac{1}{f}$ power spectrum. Hoth noise [10] or room noise is a standard background noise used for testing communication systems, as per ITU-T P.800. The power distribution of these noises over frequency can be seen in Fig. 3.

**Evaluation metrics:** The effectiveness of the proposed approach is evaluated using the following metrics:

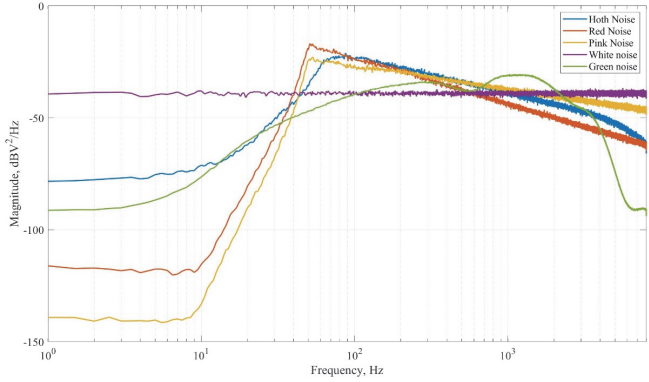

Fig. 3. Power spectrum density of noise signals used for testing.

- Signal-to-Noise Ratio Gain (SNR): measures the variation introduced by a method in the target speech to background noise ratio. More specifically, it is defined as the ratio between output SNR and input SNR, where each SNR is the ratio of target speech power over background noise power.

- Signal Interference ratio (SIR): SIR assesses the decoupling of the speech and driver sources by the MWF, i.e., it measures the variation introduced by a method in the ratio of desired speech to competing speech source(s). More specifically, it is defined as the ratio between output SIR and input SIR, where each SIR is the ratio of target speech power over the (sum of) other speech source(s) power.

- Deep Noise Suppression Mean Opinion Score (DNSMOS) [11]: DNSMOS is a deep-learning-based perceptual speech quality metric for evaluating noise suppressors. It is a non-intrusive method that estimates Mean Opinion Scores (MOS) directly from processed speech without requiring a clean reference. We use the following MOS scores predicted by DNSMOS:

   1. Speech Quality MOS – Evaluates speech clarity and quality.
   2. Background Noise MOS – Measures the extent of residual noise in speech.

SIR and SNR gain for driver and passenger speech are calculated from their respective time segments. DNSMOS is computed using both time segments with driver speech and time segments with passenger speech. For MWF, DNSMOS is calculated on the sum of the outputs from the MWF filters (for driver and passenger), and this is compared to DNSMOS computed with direct mixing of the microphone signals.

## IV. RESULTS

### A. Notch effect

To investigate the effect of MWF on the notch filtering effect, white noise is used as the directional sources (instead of speech), because it has a flat frequency spectrum. This helps to easily detect the notches in the plot. The frame sizes are varied from 100ms to 20ms to 8ms. The cross-side attenuation, which is the attenuation measured at a microphone for a source signal coming from an opposite-side source, is also varied. For example, the original simulated impulse responses have a cross-side attenuation of 2dB, which means that the driver(passenger) signal reaching the secondary(primary) microphone will be 2dB weaker than the driver(passenger) signal at the primary(secondary) microphone. To investigate whether the cross-side attenuation affects the notch filtering effect, we also considered a case where the simulated impulse responses are modified to increase the cross-side attenuation to 10dB. The results for the different cases are shown in Fig. 4

### B. Noise reduction

An 8ms frame size, with a regularization factor $\delta$ of 1.0 and a moving average factor $\lambda$ of 0.96, provided the best noise reduction across different noise types. For hoth noise, a regularization factor of 100 is used for frequencies lower than 312.5 Hz and 1.0 for higher frequencies. From Table II, we see that for the MWF the SNR gain is about 10 dB with white noise, 6 dB with red/pink noise, 9 dB with green noise, and 7 dB with hoth noise. The input SNR was also varied from 10 dB to 0 dB to see how the SNR gain and DNSMOS values vary with the noise level, with results shown in Table III for white noise. Without the MWF, i.e., simply summing (mixing) the microphone signals, resulting SNR and SIR gains of around -1dB and -1.26 dB are found.

### C. Head movement

During simulation, the original impulse responses were changed to assume 0.1 m and 0.15 m lateral displacement of the driver's head position during the 36-52 sec. interval, to see if the speaker's head movements affect MWF performance. This was done by changing the driver distance from the left from 0.6 m. to 0.5 m. and 0.45 m., with different input SNRs (Figs. 5,6). With white noise and input SNR of 5dB, the driver's SNR gain slightly decreases from 7.68 to 7.47 dB. With input SNR of 1 dB, the change is from 8.81 to 8.25 dB. There is therefore hardly any drop in SNR gain in both cases, and the MWF was able to adapt quickly to the driver's head position change. Test were also performed with adaptation paused after 36 seconds, to see how important it is for the filter to continuously adapt. The comparative results are shown in Figs. 5-7.

TABLE II. PERFORMANCE OF MWF OUTPUTS SUM FOR 5 DB INPUT SNR AND DIFFERENT NOISES. FOR MIC. SUM, SNR AND SIR GAINS ARE -1DB, -1.26 DB.

| Noise type | MWF outputs sum: SNR, SIR gains, driver (dB) | MWF outputs sum: SNR, SIR gains, passenger (dB) | MWF outputs sum: DNSMOS speech, noise | Mic. sum: DNSMOS speech, noise |
|---|---|---|---|---|
| White | 10.14, 8,28 | 10.21, 7.53 | 3.46, 2.55 | 2.94, 1.64 |
| Red | 5.77, 7.91 | 6.10, 6.96 | 3.47, 2.79 | 1.22, 1.17 |
| Pink | 6.36, 8.58 | 6.13, 8.45 | 3.33, 2.58 | 1.24, 1.17 |
| Green | 8.91, 9.30 | 8.63, 8.82 | 3.33, 2.61 | 1.88, 1.36 |
| Hoth | 7.04, 7.36 | 6.61, 8.33 | 2.82, 2.18 | 1.19, 1.16 |

TABLE III. PERFORMANCE OF MWF OUTPUTS SUM, FOR DIFFERENT INPUT SNRs, WHITE NOISE

| Input SNR | MWF outputs sum: SNR, SIR gains, driver (dB) | MWF outputs sum: SNR, SIR gains, passenger (dB) | MWF outputs sum: DNSMOS speech, noise | Mic. sum: DNSMOS speech, noise |
|---|---|---|---|---|
| 10 | 9.05, 7.32 | 9.24, 6.83 | 3.61, 2.98 | 3.41, 1.89 |
| 5 | 10.14, 8.28 | 10.21, 7.53 | 3.46, 2.55 | 2.94, 1.64 |
| 0 | 10.86, 9.96 | 10.73, 8.70 | 3.01, 1.95 | 1.21, 1.15 |

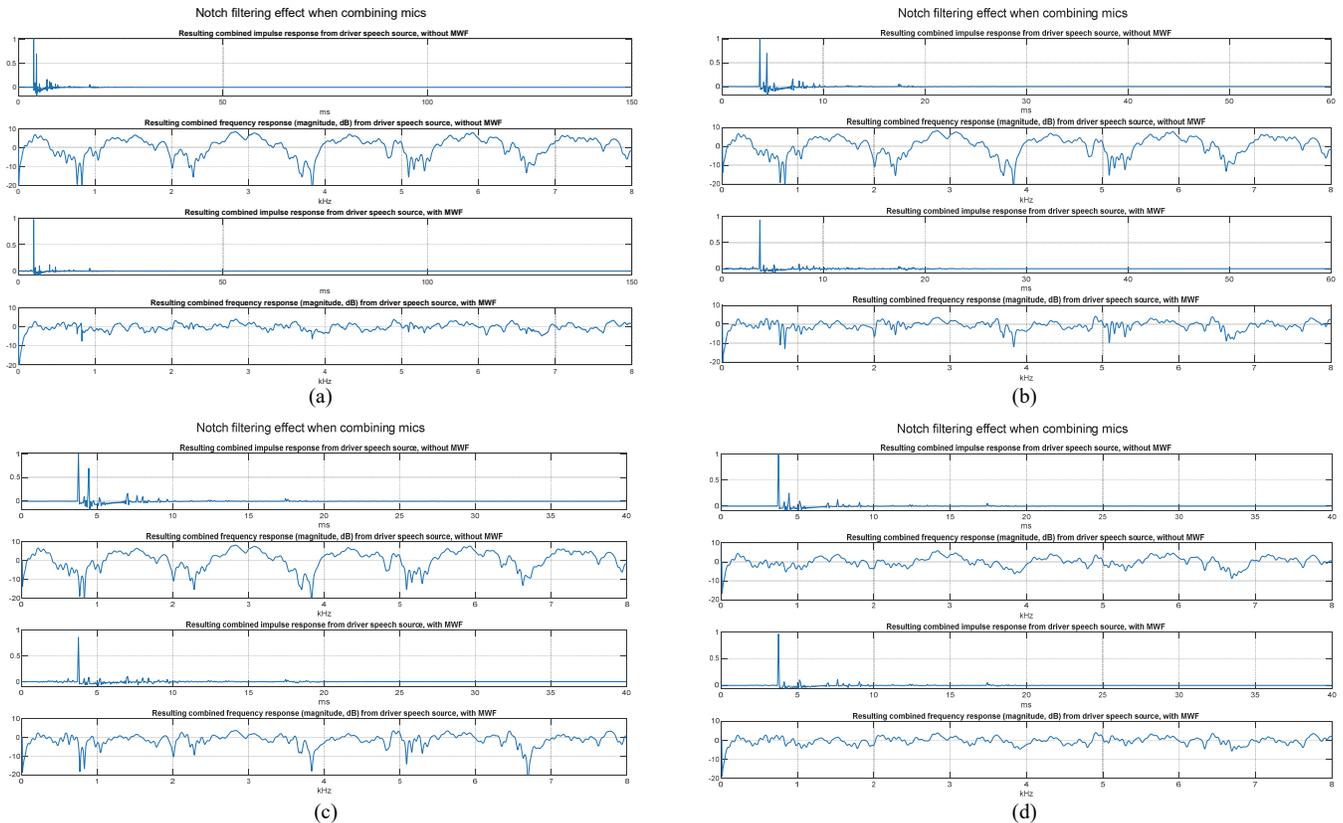

Fig. 4. Spectral notches with MWF outputs sum and without MWF outputs sum (i.e., with simple mic. signals sum), for different frame sizes and different cross-side impulse response attenuation: a)100ms, 2 dB, b) 20ms, 2 dB, c) 8ms, 2 dB, d) 8ms, 10 dB.

## V. Discussion

### A. Notch filtering effect

When white noise is used as input, compared to mixing the two microphone signals, the sum of MWF filters outputs shows a clear reduction in the notch filtering effect in Fig. 4. The notch filtering effect is more effectively removed when the frame size is higher, with almost all notches removed at 100 ms frame size. The notch filtering effect is reduced when the impulse responses are modified to have higher cross-side attenuation. When the cross-side attenuation was set to 10dB, the notch filtering effect was negligible, and the MWF frame size had less impact on the performance. When dealing with speech signals, the statistics change with time, and a smaller frame size is required for quick adaptation as opposed to directional white noise sources.

### B. Noise Reduction

SNR gain and DNSMOS help to determine the level of noise reduction. DNSMOS scores give values from 0-5, where a signal score closer to 5 denotes better speech quality, and a noise score closer to 5 denotes better noise reduction. In every case, with MWF the SNR gain and DNSMOS values, especially DNSMOS noise scores, are higher than when the microphone signals are mixed. The best performance is seen in the case of white noise, with an SNR gain of about 10dB. DNSMOS signal scores also show an increase with MWF. Hoth noise, commonly used for testing in communication systems, also shows an improvement when MWF is used, both in signal and noise scores.

### C. Head Movement

When the speaker moved his head during the speech, there was hardly any drop in performance, as long as there was continuous adaptation with the filter coefficients being updated every 8ms. Regardless of input SNR, if the adaptation is paused the MWF struggles to maintain its original performance even when the head of the speaker is back to the original position (interval 52-68 sec.). Figs. 5-7 were produced with white noise but other noises also showed similar behavior. If we compare the performance between the cases where the signals are directly mixed and the cases where MWF is present, the DNSMOS values indicate the importance of continuous adaptation.

## VI. Conclusion

In this paper, the drawbacks of simple mixing (adding) and switching methods to combine two microphone signals into a single signal were first explained, i.e., the notch filtering effects and increased noise in combined signal, and decreased AEC performance when switching two microphone signals before a single AEC. To resolve these issues, we investigated the use of an adaptive MWF filter to separate driver and passenger signals before combining them, thereby reducing notch filtering effects and minimizing background noise.

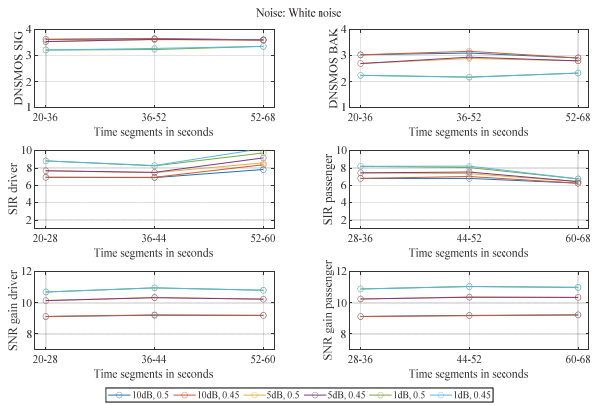

Fig. 5. Performance scores for MWF outputs sum with head movement for different input SNRs and driver movements, continuous MWF adaptation.

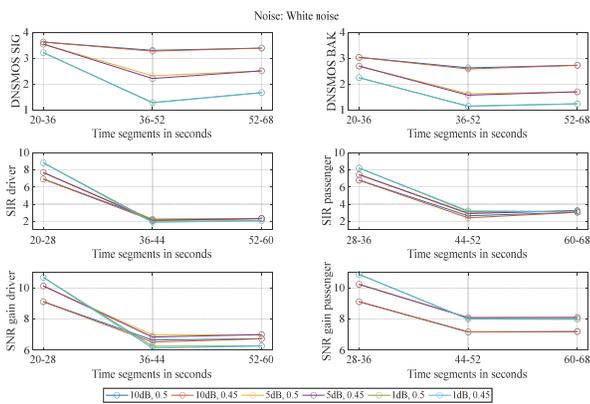

Fig. 6. Performance scores for MWF outputs sum with head movement for different input SNRs and driver movements, MWF adaptation paused after 36 seconds.

The MWF proved effective in a car environment, where each front microphone is mostly dedicated to a speaker, which makes the task of estimating the signal correlations easier (i.e., it becomes easier to detect when each source is active and to estimate the related correlations). It successfully removed notch filtering effects, particularly with larger frame sizes. Through multichannel linear filtering, which is nearly LTI, the MWF effectively reduced background noise with minimal distortion. Since the number of acoustic sources (talkers or loudspeakers) that MWF filters can extract is limited by the number of microphones, MWF can be disabled in the presence of loudspeaker signals, or, alternatively and at higher cost, AEC could be performed on each microphone signal before the signals are processed by the MWFs. The MWF filters can perform driver and passenger speech extraction whether only one of them is talking or whether they are talking simultaneously. In the latter case, since the MWF extracts the speech sources before combining them, it could also be used for interference cancellation (i.e., extracting one speech source and cancelling competing talkers).

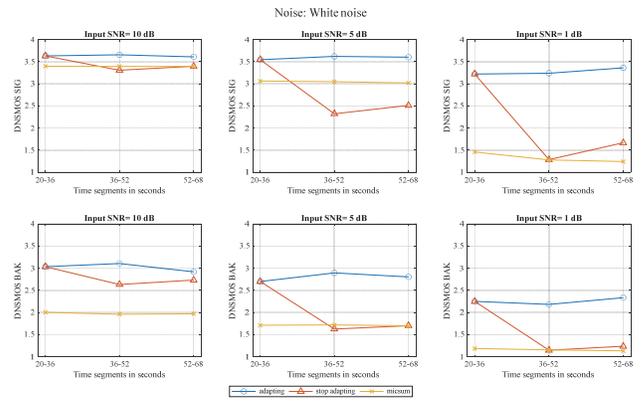

Fig. 7. DNSMOS performance scores for MWF outputs sum and simple microphone signals sum with head movement.

Talker movements could potentially impact MWF performance, but in such cases the MWF has shown to adapt quickly without divergence. The performance of MWF filtering could also be limited by highly non-stationary background noise, with time-varying spectra and time-varying signal level imbalance between microphones.

REFERENCES

[1] J. Meyer and K. U. Simmer, "Multi-channel speech enhancement in a car environment using Wiener filtering and spectral subtraction," in *Proc. IEEE International Conference on Acoustics, Speech, and Signal Processing (ICASSP)*, Munich, Germany, 1997, pp. 1167-1170.

[2] C. Fox, G. Vitte, M. Charbit, J. Prado, R. Badeau and B. David, "A subband hybrid beamforming for in-car speech enhancement," in *Proc. 20th European Signal Processing Conference (EUSIPCO 2012)*, Bucharest, Romania, 2012, pp. 11-15.

[3] S. Stenzel and J. Freudenberger, "On the Speech Distortion Weighted Multichannel Wiener Filter for diffuse Noise," in *Proc. 10th ITG Symposium on Speech Communication,* Braunschweig, Germany, 2012, pp. 1-4.

[4] M. Gimm, F. Kühne and G. Schmidt, "A Multichannel Spatial Hands-Free Application for In-Car Communication Systems," in *Towards Human-Vehicle Harmonization*, Germany: De Gruyter, 2023, ch. 10, pp.129-140.

[5] B. Kaulen, J. Abshagen, and G. Schmidt, "Multichannel Wiener filter in active sound-navigation-and-ranging systems - A joint beamformer and matched filter approach, " *IET Radar, Sonar & Navigation*, vol. 18, no. 9, pp.1554-1569, Sept. 2024.

[6] S. Gannot and I. Cohen, "Adaptive Beamforming and Postfiltering," in *Springer Handbook of Speech Processing*, Germany: Springer, 2008, ch. 47, pp. 945-978.

[7] T. Van den Bogaert, S. Doclo, J. Wouters and M. Moonen, "Speech enhancement with multichannel Wiener filter techniques in multimicrophone binaural hearing aids," *The Journal of the Acoustical Society of America*, vol. 125, no.1, pp. 360-371, Jan. 2009.

[8] RIR-Generator [Online]. Available: https://github.com/ehabets/RIR-Generator.

[9] J.B. Allen and D.A. Berkley, "Image method for efficiently simulating small-room acoustics," *The Journal of the Acoustical Society of America*, vol. 65, no. 4, pp. 943-950, April 1979.

[10] D.F. Hoth, "Room noise spectra at subscribers' telephone locations," *The Journal of the Acoustical Society of America*, vol. 12, no. 3, pp. 499-504, Apr. 1941.

[11] C. K. A. Reddy, V. Gopal, and R. Cutler, "DNSMOS P.835: A Non-Intrusive Perceptual Objective Speech Quality Metric to Evaluate Noise Suppressors," *arXiv preprint arXiv:2110.01763*, pp.1-5, Feb. 2022. [Online]. Available: https://arxiv.org/abs/2110.01763